\documentclass{elsart}

\usepackage{natbib}

\usepackage{graphicx}
\usepackage{amssymb}

\begin{document}

\begin{frontmatter}

% Title, authors and addresses

% use the thanksref command within \title, \author or \address for footnotes;
% use the corauthref command within \author for corresponding author footnotes;
% use the ead command for the email address,
% and the form \ead[url] for the home page:
% \title{Title\thanksref{label1}}
% \thanks[label1]{}
% \author{Name\corauthref{cor1}\thanksref{label2}}
% \ead{email address}
% \ead[url]{home page}
% \thanks[label2]{}
% \corauth[cor1]{}
% \address{Address\thanksref{label3}}
% \thanks[label3]{}

\title{The search for milky way halo substructure WIMP annihilations using the GLAST LAT}

% use optional labels to link authors explicitly to addresses:
% \author[label1,label2]{}
% \address[label1]{}
% \address[label2]{}

\author{Larry Wai}

\address{Stanford Linear Accelerator Center, Stanford, California \\ {~} \\
{\em representing the GLAST LAT Collaboration} \\ {\em Dark
Matter and New Physics Working Group}}

\begin{abstract}
% Text of abstract
The GLAST LAT Collaboration is one among several experimental groups,
covering a wide range of approaches, pursuing the search for the
nature of dark matter.  The GLAST LAT has the unique ability to find
new sources of high energy gamma radiation emanating directly from
WIMP annihilations in situ in the universe.  Using it's wide band
spectral and full sky spatial capabilities, the GLAST LAT can form
``images'' in high energy gamma-rays of dark matter substructures in
the gamma-ray sky.  We describe a preliminary feasibility study for
indirect detection of milky way dark matter satellites using the GLAST
LAT.
\end{abstract}


\end{frontmatter}

% main text

%\section{Introduction}

Particle dark matter in the form of as yet to be discovered weakly
interacting massive particles (WIMPs) can be discovered by direct
detection (WIMP-nucleus scattering), indirect detection (WIMP pair
annihilation), or collider production (WIMP pair production).  Direct
detection depends upon the local density of WIMP particles, whereas
indirect detection depends upon the density (squared) of WIMP
particles in regions of the galaxy or beyond.  Indirect detection also
depends upon WIMPs being either Majorana particles or a mixture of
roughly equal numbers of WIMPs and anti-WIMPs.  Indirect detection
experiments can be categorized by particle type which include
neutrinos, anti-matter, and gamma-rays \cite{gammas}.  Here we focus
on how the Gamma Ray Large Area Space Telescope (GLAST) Large Area
Telescope (LAT) detector \cite{glast} may observe the particle nature
of dark matter.
%
%The GLAST LAT consists of a tracking detector, calorimeter, and
%anti-coincidence detector.  The tracking detector is composed of 36
%layers of tungsten and silicon strip detectors, and serves as the
%target for converting the gamma rays as well as measuring the tracks
%of the conversion electron-positron pair.  The calorimeter is composed
%of 8 layers of stacked Cesium Iodide crystal logs readout with PIN
%photodiodes at both ends of each log.  The anti-coincidence detector
%is composed of overlapping plastic scintillator tiles readout with
%PMTs via wavelength shifting fibers.  At the time of writing of this
%paper, the entire detector is fully assembled, and the testing of the
%detector in a space-like environmental test is in full swing.  The
%launch of GLAST into low earth orbit is on schedule for September
%2007.
%
%The LAT is designed to present an effective area of $\sim$10000 square
%centimeters with a field of view covering $\sim$1/6 of the full sky.
%The standard mode of operation will be a continuous scan of the sky.
%The energy threshold is 20 MeV and gamma ray energies up to 300 GeV
%can be measured.  The angular resolution for the LAT is significantly
%better than that of EGRET (factor of at least a few), as is the energy
%resolution, especially above 10 GeV.  The mission lifetime is designed
%for 5 years with a goal of 10 years.  The resulting point source
%sensitivity ($\sim$$3\times 10^{-9}cm^{-2}s^{-1}$) is $\sim$30 times
%better than that of EGRET.
%
%\section{Number of observable dark matter sources}

\begin{figure}
\begin{center}
\includegraphics[width=1.0\textwidth]{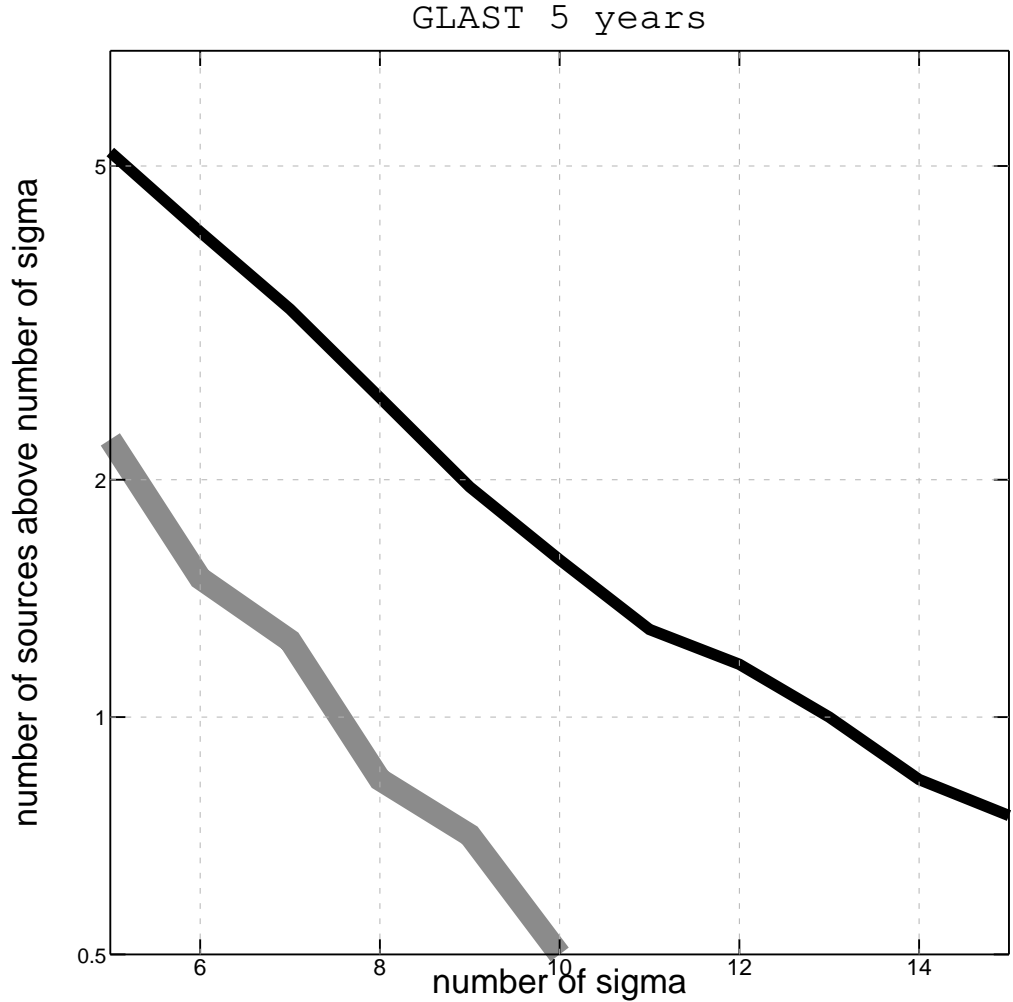}
\caption{This graph shows the number of dark matter sources observed
in 5-years of GLAST data above a given significance in number of
sigma.  The dark matter sub-halo distribution for $>10^6$ solar masses
was estimated with the semi-analytic method of Taylor \& Babul
\cite{TaylorBabul}.  The thin black line corresponds to the SUSY model
LCC2 and the thick grey line corresponds to the SUSY model LCC4, as
defined in Baltz et.al. \cite{Baltz}.  The diffuse gamma ray
background is estimated using the $>$1~GeV source subtracted map from
Cillis \& Hartman \cite{CillisHartman}.\label{nclump_vs_nsigma-fig}}
\end{center}
\end{figure}

The standard model of cosmology relies on a hierarchical distribution
of dark matter which on galactic scales implies a significant number
of as yet unobserved dark matter satellites with masses less than
approximately $10^7$ solar masses, the mass scale of the known dark
matter dominated low surface brightness dwarf galaxies.  We have
estimated the number of Milky Way dark matter satellites with $>10^6$
solar masses observable by GLAST.  The dark matter sub-halo
distribution was estimated with the semi-analytic method of Taylor \&
Babul \cite{TaylorBabul}.  The dark matter satellite distribution is
roughly spherically symmetric about the galactic center and extends
well beyond the solar orbit; thus the dark matter satellites are
located mostly at high galactic latitudes.  We assumed that the dark
matter satellites emit gamma rays via the decay of neutral pions.  The
neutral pions are produced in hadronization of final state quarks, tau
leptons, or W/Z bosons produced by pair annihilation of WIMPs.  As a
case study, we calculated the gamma ray flux for the LSP WIMP (SUSY)
for the two benchmark points LCC2 and LCC4 as defined in Baltz,
et.al. \cite{Baltz}. The background was estimated using the EGRET
point source subtracted sky map above 1GeV from Cillis \& Hartman
\cite{CillisHartman}.  The significance of the dark matter signal was
then estimated to be the number of signal events within the satellite
tidal radius (or the PSF 68\% containment radius, whichever was
bigger) divided by the square root of the number of background events
within the same radius.  The resulting number of dark matter
satellites with significance of at least some number of sigma is shown
in figure~\ref{nclump_vs_nsigma-fig}.  For these benchmark SUSY
points, we would expect to observe in the range of a few dark matter
satellites with 5 years of GLAST data.

%\section{Dark matter source identification}

\begin{figure}
\begin{center}
\includegraphics[width=1.0\textwidth]{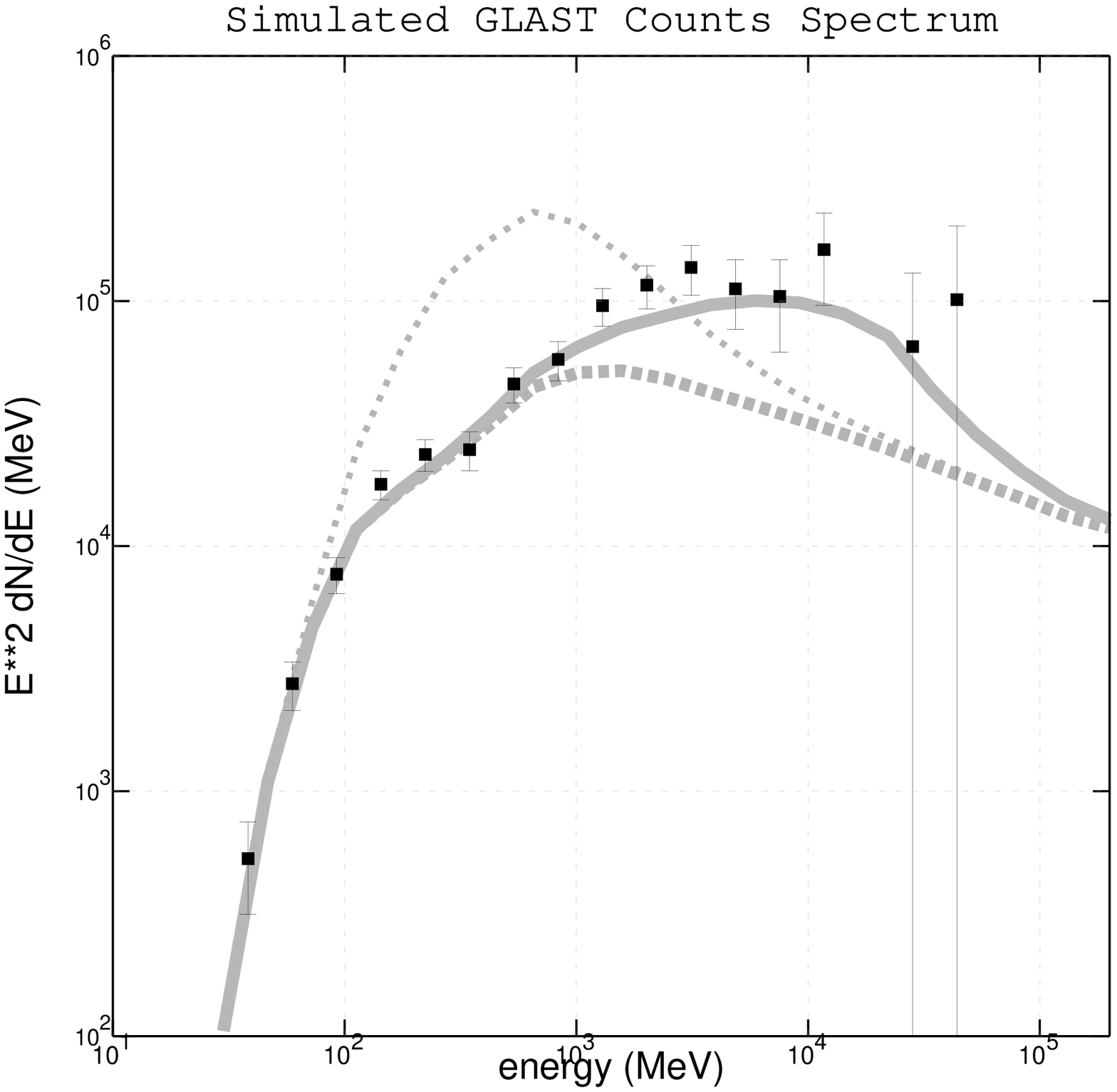}
\caption{This graph shows the number of counts versus energy for a
simulated 55-days of GLAST.  Simulated data from the detector is shown
as squares with error bars; this is composed of diffuse background and
a 100GeV mass WIMP (approximately 10-sigma detection).  The solid grey
line is the fitted counts spectrum, and the thick dotted grey line is
the diffuse background only, i.e. optimized background model from
Strong, Moskalenko, Reimer \cite{galprop} plus an isotropic diffuse
background from Sreekumar et. al. \cite{Sreekumar}.  The thin dotted
grey line is the diffuse background plus a source with a simple
powerlaw spectrum, with spectral index
-2.6.\label{counts_vs_energy-fig}}
\end{center}
\end{figure}

In order to distinguish a dark matter source from a more typical
astrophysical source, such as a pulsar or molecular cloud, we can make
use of the GLAST spatial and spectral resolving capabilities.  The
typical extent of the observable dark matter satellites is in the
range of 1 degree.  GLAST should be able to resolve these sources
using the WIMP annihilation photons above 1 GeV, for which the PSF is
approximately 1/2 degree.  The WIMP pair annihilation spectrum is
extremely hard, much harder than any of the EGRET sources, except
possibly the galactic center source.  Molecular cloud gamma ray
sources will typically have a power law energy distribution, which is
distinct from the WIMP annihilation spectrum.  In
figure~\ref{counts_vs_energy-fig} we plot the counts spectra for a 100
GeV mass WIMP, high latitude ($b=-31$deg), 10-sigma source, plus
diffuse background.  The diffuse background model consists of the
optimized background model from Strong, Moskalenko, Reimer
\cite{galprop} plus the isotropic diffuse background described in
Sreekumar et. al. \cite{Sreekumar}.  Identification of high latitude
GLAST sources with energy spectra consistent with WIMP annihilation
would provide a excellent targets for imaging atmospheric Cherenkov
telescope (IACT) observations.  Precision high statistics observations
of WIMP sources can yield information on the nature of the WIMP.  For
example, the measurement of the $\gamma\gamma$ and/or $\gamma Z^0$
line branching fraction can yield information complementary to that
obtained with accelerator or direct detection experiments.

%\label{}

% The Appendices part is started with the command \appendix;
% appendix sections are then done as normal sections
% \appendix

% \section{}
% \label{}

% Bibliographic references with the natbib package:
% Parenthetical: \citep{Bai92} produces (Bailyn 1992).
% Textual: \citet{Bai95} produces Bailyn et al. (1995).
% An affix and part of a reference:
%   \citep[e.g.][Ch. 2]{Bar76}
%   produces (e.g. Barnes et al. 1976, Ch. 2).

%\begin{thebibliography}{}

% \bibitem[Names(Year)]{label} or \bibitem[Names(Year)Long names]{label}.
% (\harvarditem{Name}{Year}{label} is also supported.)
% Text of bibliographic item

%\bibitem[]{}

%\end{thebibliography}

\end{document}